\documentclass[fleqn,usenatbib,useAMS]{mnras}
\usepackage[T1]{fontenc}
\DeclareRobustCommand{\VAN}[3]{#2}
\let\VANthebibliography\thebibliography
\def\thebibliography{\DeclareRobustCommand{\VAN}[3]{##3}\VANthebibliography}
\usepackage{graphicx}
\usepackage{amsmath}
\usepackage{bm}
\usepackage[version=4]{mhchem}
\usepackage{upgreek}
\usepackage{hyperref}
\usepackage{float}
\usepackage{siunitx}
\usepackage{textcomp} 
\usepackage{float}
\usepackage{capt-of}
\usepackage[dvipsnames]{xcolor}
\usepackage{makecell}

\title[Computing Anharmonic IR Spectra of PAHs Using MLMD]{Computing Anharmonic Infrared Spectra of Polycyclic Aromatic Hydrocarbons Using Machine Learning Molecular Dynamics}

\author[X. Mai et al.]{Xinghong Mai$^{1}$, Zhao Wang$^{1}$ \thanks{E-mail: zw@gxu.edu.cn (ZW)}, Lijun Pan$^{1}$, Johannes Sch\"orghuber$^{2}$, P\'eter Kov\'acs$^{2}$,\newauthor
Jes\'us Carrete$^{3}$, and Georg K. H. Madsen$^{2}$
\\
$^{1}$Laboratory for Relativistic Astrophysics, Department of Physics, Guangxi University, 530004 Nanning, China\\
$^{2}$Institute of Materials Chemistry, TU Wien, 1060 Vienna, Austria\\
$^{3}$Instituto de Nanociencia y Materiales de Aragón (INMA), CSIC-Universidad de Zaragoza, E-50009 Zaragoza, Spain\\
}

\pubyear{2025}

\usepackage{newtxtext}
\begin{document}
\label{firstpage}
\pagerange{\pageref{firstpage}--\pageref{lastpage}}
\maketitle

\begin{abstract}
We introduce a machine learning molecular dynamics (MLMD) approach to calculate the anharmonic infrared (IR) absorption spectra of polycyclic aromatic hydrocarbons (PAHs), key carriers of interstellar aromatic IR bands. This method accounts for temperature effects in a molecule-specific way and achieves accuracy comparable to conventional quantum chemical calculations at a fraction of the cost, scaling linearly with system size. We applied MLMD to calculate the anharmonic spectra of 1,704 PAHs in the NASA Ames PAH IR Spectroscopic Database with up to 216 carbon atoms at different temperatures, demonstrating its capability for high-throughput spectral calculations of large molecular systems.\\
\end{abstract}


\section{Introduction}

In the mid-1970s, distinct infrared (IR) signals were identified in the universe, notably at wavelengths of 3.3, 6.2, 7.7, 8.6, 11.2, and 12.7 $\upmu$m. These signals were later linked to the vibrations of C-H and C-C bonds in aromatic molecules, known as aromatic IR bands (AIB) \citep{leger1984,allamandola1985}. Such emissions are primarily attributed to polycyclic aromatic hydrocarbons (PAHs) after the absorption of ultraviolet (UV) photons. PAHs are recognized as key contributors to the evolution of the interstellar medium, influencing processes such as gas heating, ionization balance, and star formation \citep{Tielens2008}. Determining the population, composition, size distribution, charge state, and chemical structures of interstellar PAHs provides valuable insight into the physical conditions of their host galaxies. However, this remains a significant challenge due to the structural diversity of PAHs. Variations in carbon-ring arrangements, side groups, substitutions, charge states, and isotopologues lead to distinct IR emission characteristics, each governed by specific vibration modes \citep{peeters2021}.

In order to tackle these challenges, quantum chemical calculations (QCCs) are now an indispensable aid in AIB analysis. Using QCCs, publicly available spectral databases have been established for AIB analysis, such as the NASA Ames PAH IR Spectroscopic Database (PAHdb) \citep{mattioda2020,Bauschlicher2018,boersma2014}. The availability of these databases enables extensive data-driven analysis of AIBs, providing a more robust interpretation of their features, see, e.g., \citep{boersma2013,boersma2015,Maragkoudakis2020,sadjadi2015b,cami2011,shannon2019,ricca2021}. Furthermore, these databases have facilitated the analysis of astrochemical components using machine learning (ML) models, demonstrating significant computational efficiency \citep{Kovacs2020} and helping to uncover relationships between molecular structures and spectra \citep{Meng2021,Meng2023}. Despite these advancements, current theoretical spectral databases face three primary limitations: 

\begin{itemize}
\item \textbf{Harmonic oscillator approximations}: Most spectra neglect anharmonic vibrations, leading to significant discrepancies with experimental observations \citep{mackie2015,mackie2018a}.

\item \textbf{Neglect of molecule-specific thermal effects}: Most theoretical spectra in databases such as PAHdb are calculated at the ground state, without accounting for temperature effects. Consequently, emission features derived from these absorption spectra involve significant approximations.

\item \textbf{Underrepresentation of large PAHs}: Although PAHs with 50 or more C atoms are considered more likely AIB carriers \citep{Sellgren1984, Allamandola1989}, they are significantly underrepresented in current databases.
\end{itemize}

These challenges arise mainly from the high computational cost of QCCs. In particular, the computation time for anharmonic molecular spectra using second-order vibrational perturbation theory (VPT2) is prohibitively long, typically only accommodating molecules with $N_\text{C}<25$, where $N_\text{C}$ represents the number of carbon atoms in the system \citep{mackie2016,esposito2024a,maltseva2016,maltseva2018,lemmens2019}. The cost of incorporating temperature effects into these calculations using the Wang-Landau random walk algorithm is even more prohibitive, being orders of magnitude higher due to the need for extensive sampling \citep{chen2018a,chen2019}. Facing the vast diversity of PAH species, these limitations significantly hinder the accurate interpretation of observational data, posing substantial challenges in the era of advanced observational facilities such as the James Webb Space Telescope (JWST) \citep{boersma2023,ricca2024}.

As a result, there is an urgent need for a cost-effective and accurate method that incorporates both anharmonic and temperature effects in the construction of spectral databases to improve our understanding of the origin of AIBs. In this work, we propose a ML-based molecular dynamics (MLMD) approach for computing PAH IR absorption spectra. This method replaces computationally expensive electronic structure calculations with far more efficient ML-based ones. More importantly, it explicitly accounts for anharmonic vibrations and temperature effects. We show that the MLMD approach can reproduce experimental spectra with accuracy comparable to state-of-the-art VPT2 QCCs, while significantly reducing computational time, achieving a scaling approximately linear in $N_\text{C}$.

\section{Methods} 
Molecular dynamics (MD) is a powerful computational method that simulates the time-dependent behavior of atomic and molecular systems at a specified temperature by integrating Newton's equations of motion. This technique uses a potential energy surface (PES) to characterize interatomic interactions and energy changes within the system. When calculating the vibrational spectra of molecules, MD presents significant advantages by explicitly accounting for anharmonic effects, such as band combination, overtones, and mode coupling. This comes at the cost of not capturing some quantum effects whenever normal modes are likely to be found near their ground state. Beyond this fundamental consideration, classical MD has historically faced technical challenges related to accuracy and transferability. The empirical force fields used to construct the PES are often parameterized for specific molecules, which may hinder their performance when applied to others. Furthermore, although polarizable models exist \citep{bedrov2019}, classical MD generally does not account for the distribution of electrons and, therefore, cannot inherently deal with dipole moments. To address these challenges, \textit{ab initio} molecular dynamics (AIMD) can be used, where the motion of nuclei is described classically but the force contributions from the electrons are calculated quantum mechanically. However, like the QCC-based VPT2 method, AIMD is computationally expensive, imposing significant constraints on the maximum size of the systems studied.

A solution is to substitute the majority of electronic structure calculations in AIMD with more cost-effective ML computations. Using data points derived from QCCs, ML models can be trained to construct the PES and predict the charge distribution across various molecular configurations. This MLMD methodology has been proven to be highly efficient in previous studies focusing on vibration of diverse molecular types \citep{gastegger2017, zhou2021, du2024,Xu2024,Schmiedmayer2024}.

\begin{figure}
\centerline{\includegraphics[width=0.50\textwidth]{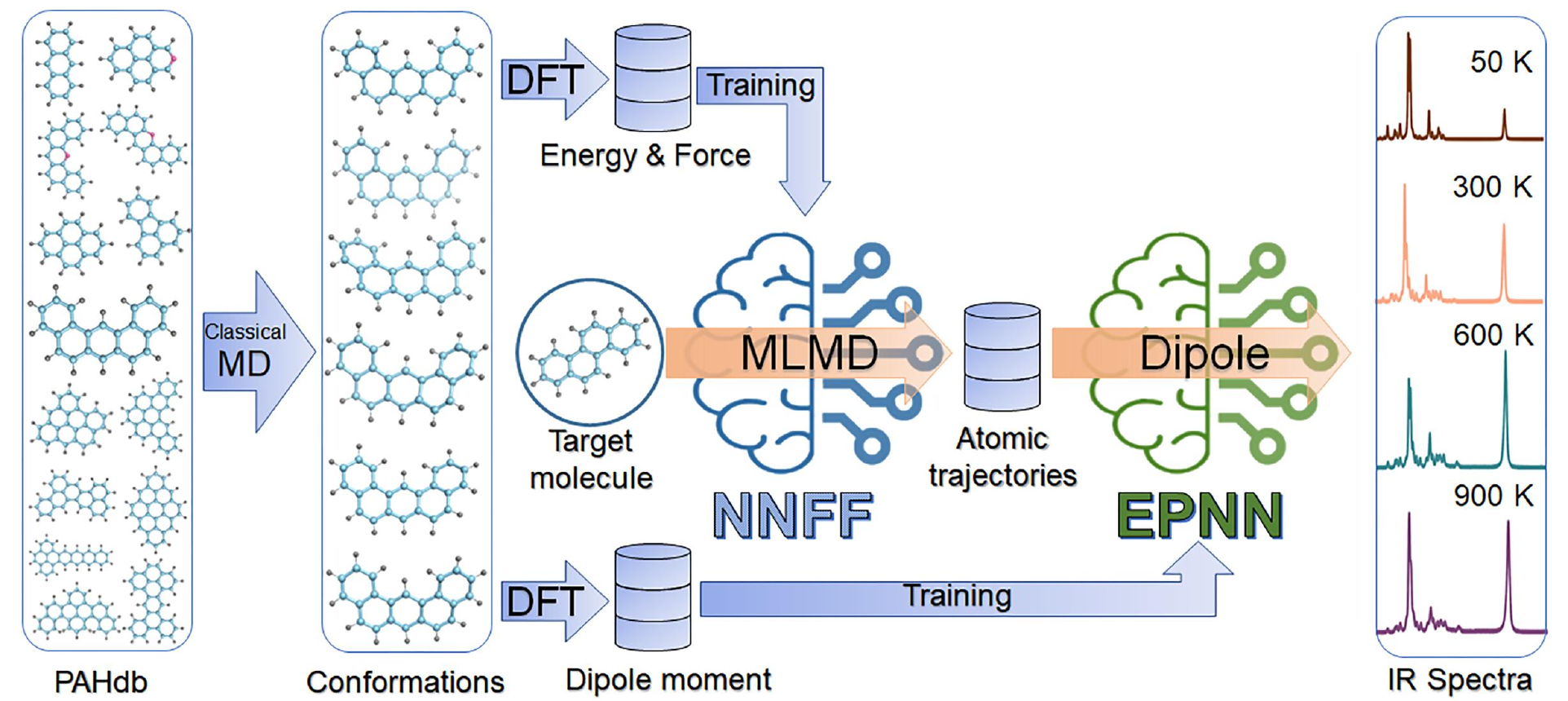}}
\caption{Schematic representation of the workflow for computing anharmonic IR absorption spectra.}
\label{F1}
\end{figure}

In this study, we build two distinct ML models: a neural-network force field (NNFF) for the construction of PES, and an electron-passing neural network (EPNN) model to predict the molecular dipole moment $\mathbf{p}$. As depicted in Figure \ref{F1}, our workflow for computing the IR spectra consists of two main phases. The first is a training phase (blue arrows) that includes:

\begin{enumerate}
\item Classical MD simulations of PAHs to generate conformations;
\item QCCs to obtain their energies, forces, and dipole moments;
\item Training of the NNFF and EPNN models using these data.
\end{enumerate}
The second phase is prediction (orange arrows), including:
\begin{enumerate}
\item MLMD to obtain conformations of vibrating PAHs via NNFF;
\item Prediction of $\mathbf{p}$ for every conformation using EPNN;
\item Spectrum computation by evaluating the time evolution of $\mathbf{p}$.
\end{enumerate}

The predicted $\mathbf{p}$ is not used as an input for the calculation of energies and forces and therefore does not determine the trajectories, since the interatomic interactions determined by the charge density distribution are already accounted for by the NNFF. A detailed description of the components is provided in the next subsection.

\subsection{Training Data} 
The chemical structures of 687 neutral PAH molecules were extracted from the theoretical dataset of PAHdb (v3.2, \citet{mattioda2020}) to construct the training datasets. The selected PAHs encompass a diverse array of chemical structures that are of astronomical significance. Classical MD simulations were performed using a custom code to generate random configurations of the atoms in each molecule at 300 K, using the adaptive interatomic reactive empirical bond order potential \citep{Stuart2000}. In these simulations, the molecules reached thermal equilibrium in 200 ps using a canonical Nos\'e-Hoover thermostat, with a time step of 0.5 fs. Ten extended structures were extracted from the atomic trajectories of each molecule. As these geometries do not necessarily fall near equilibrium configurations, random Gaussian perturbations were applied to the ground-state atomic coordinates to create 15 more configurations for each molecule to improve the generalizability of the model. Finally, in total, 17,175 configurations (also called molecular conformations) were generated. Our training dataset includes 350 molecular conformations of superhydrogenated PAHs, out of a total of 17,175 conformations.

Subsequently, single-point electronic structure QCCs were performed for each of these 17,175 configurations to calculate the potential energy, the forces in the nuclei, and the dipole $\mathbf{p}$, within the framework of density functional theory (DFT) at the B3LYP/4-31G level, as implemented in the Gaussian 16 software package \citep{frisch2016}. This level of theory offers a favorable balance between accuracy and computational efficiency for large PAHs \citep{stephens1994}, given its extensive application in PAH IR spectrum studies \citep{bauschlicher2000, ricca2012}. We note that the basis set does not include polarization functions, which may introduce more uncertainties than larger basis functions (e.g 6-31g*) for PAHs containing nitrogen \citep{ricca2021,ricca2024}. However, this choice was made to ensure computational efficiency, which is crucial for large-scale sampling. Through these calculations, we generated two datasets for training the NNFF and EPNN models, respectively. The first dataset comprises 17\,175 energy and 2\,274\,825 force data points, while the second dataset includes 17\,175 molecular dipole moments. 

\subsection{NNFF}
\begin{figure}
\centerline{\includegraphics[width=9cm]{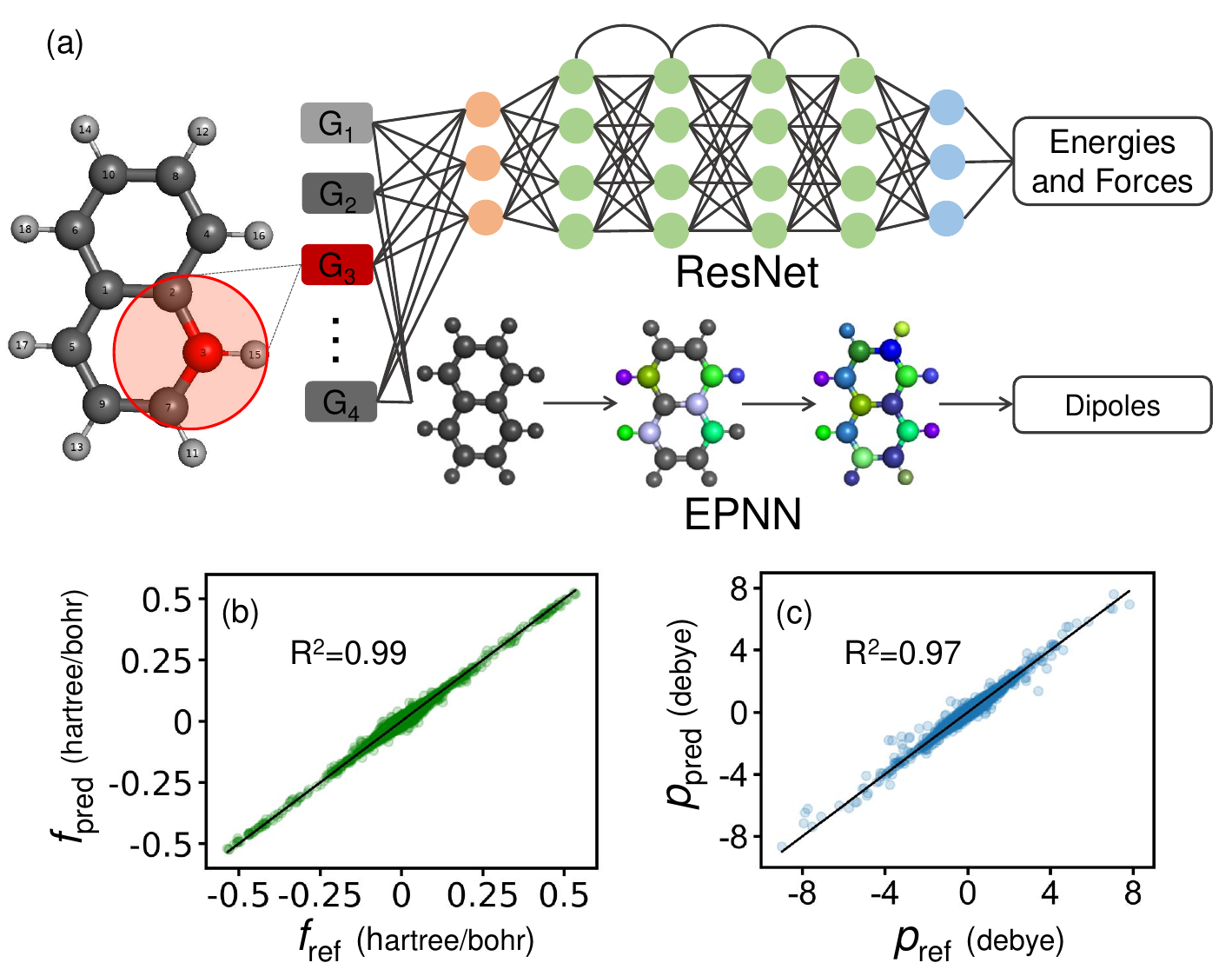}}
\caption{(a) Schematic of the operating principles of each kind of neural network (NNFF and EPNN) during inference. (b) NNFF-predicted atomic forces and (c) EPNN-predicted dipole moments vs. the DFT-calculated reference values for 3\,580 tested PAH configurations.}
\label{F2}
\end{figure}

The NNFF used in this study is based on NeuralIL \citep{montes-campos2021, carrete2023}, a refinement of the descriptor-based template introduced by Behler \citep{behler2007}. NeuralIL employs spherical Bessel descriptors \citep{kocer2020} to encode the atomic configurations in a translationally and rotationally invariant fashion, as illustrated in Figure \ref{F2}a. We set the cutoff radius of the descriptors to 3.8 \AA, with a maximum radial order of 6, ensuring effective capture of local geometric information and sufficient resolution to describe atomic interactions within the molecule. 

NNFF utilizes a deep residual network architecture (ResNet) \citep{he2016a}, implemented within the JAX framework \citep{bradbury2018}. A ResNet employs skip connections to learn the residual function between the input and output layers, effectively addressing the data degradation problem that arises with increased network depth \citep{he2016b}. This feature is particularly beneficial for modeling complex molecules such as large PAHs, as it enables the capture of deep dependencies and interaction patterns between atoms through multilevel learning, which ultimately leads to more accurate predictions of physicochemical properties \citep{zhang2024,xue2024}.

In this study, the NNFF is trained based on the previously mentioned DFT-calculated dataset of energies $E$ and atomic forces $\mathbf{f}$, with the loss function $L$ defined as a weighted sum of the differences between the predicted and reference values:

\begin{flalign}
\label{eq1}
&&
\begin{split}
L &=  0.99 \sum^{N_{\text{mol}}} \left \langle   \frac{0.2}{N_\text{atom}} \sum_{}^{N_\text{atom}} \mathrm{ ln} \left [ \mathrm{ cosh} \left (  \frac{\left \| \mathbf{f}_\text{pred}-\mathbf{f}_\text{ref} \right \|_{2}}{0.2}\right ) \right ] \right \rangle \\ 
& +0.01 \sum^{N_{\text{mol}}} \left \langle   0.02\ \mathrm{ ln} \left [   \mathrm{ cosh} \left (  \frac{E_\text{pred}-E_\text{ref}}{N_\text{atom}\times 0.02}\right ) \right ] \right \rangle,
\end{split}
&&
\end{flalign}

\noindent where $N_{\text{mol}}$ denotes the total number of molecules, and ${N_\text{atom}}$ represents the total number of atoms in a molecule. The subscripts $_\text{pred}$ and $_\text{ref}$ denote the values predicted by NNFF and DFT, respectively. $\left \| \mathbf{f}_\text{pred}-\mathbf{f}_\text{ref} \right \|_{2}$ represents the Euclidean norm of the vector difference between the predicted force and the reference force. The model adopts a core-width sequence of 64:32:16:16 and employs the fully non-linear VeLO optimization technique \citep{metz2022} for 600 epochs of training to enhance convergence \citep{carrete2023}.

To evaluate the performance of the model, we conducted tests on a set of configurations for 3\,580 PAHs that were excluded from the training set. The results presented in Figure \ref{F2}b demonstrate that the NNFF model accurately reproduces the forces calculated by DFT, achieving a coefficient of determination (${R^2}$) of approximately 0.99 and a mean root mean square error (RMSE) of 0.00117 hartree/bohr (see Table \ref{T1}), indicating strong generalizability to unseen samples. This capability allows the NNFF model to effectively replace DFT electronic structure calculations in AIMD, providing robust support for subsequent IR spectrum calculations.

\begin{table}
\caption{\label{T1} Average RMSE of the predicted energy and force from the NNFF, and of the predicted molecular dipole moments from the EPNN, for samples in the training, validation and test datasets.}
\begin{center}
\begin{tabular}{l c c c}
\hline
  & Train & Validation & Test \\
\hline
Energy (hartree per atom) & 0.315 & 0.323 & 0.319 \\
Force (hartree/bohr) & 0.00118 & 0.00117 & 0.00115 \\
Dipole (debye) & 0.0550 & 0.0545 & 0.0556 \\
\hline
\end{tabular}
\end{center}
\end{table}

The training data of NNFF were computed at the B3LYP/4-31G level of DFT, which is known to systematically overestimate the bond strength. To correct for this, we applied a scaling factor of 0.9578 to the MLMD-predicted vibrational frequencies, following the approach proposed by \citet{bauschlicher1997}.

\subsection{EPNN}
Predicting the molecular dipole moment $\mathbf{p}$ in a picture of discrete atoms presents a significant challenge due to its dependence on the distribution of atomic partial charges (PC), for which no unique determination method exists. Various PC partitioning schemes can produce different $\mathbf{p}$ for the same molecule. To avoid this issue, we directly train neural networks to predict $\mathbf{p}$, bypassing the need to explicitly calculate PCs as training data, following the approach of \citet{gastegger2017}. However, we utilize a different neural network architecture for enhanced predictive ability. Specifically, we adopt a message-passing neural network known as the EPNN \citep{metcalf2021} that guarantees the conservation of charge. In order to achieve charge conservation, the EPNN adopts a graph neural network architecture and ensures anti-symmetry of the updates with respect to permutation of the input indices. In the present implementation of the EPNN the node states are extended by concatenation with the spherical Bessel descriptor for each individual atom. Charge conservation is preserved as these descriptors are not modified in the update phase.

The EPNN was trained over 400 epochs using VeLO, employing the aforementioned DFT-calculated $\mathbf{p}$ dataset. The loss function used for training is defined as

\begin{flalign}
\label{eq2}
&&
L = \sum\limits_{\alpha=1}^{3} \sum_{i=1}^{N_{\text{mol}}}\ln \left [ \cosh \left (\frac{{ {p}^{\left(\alpha\right)}_{i,\text{pred}}- {p}^{\left(\alpha\right)}_{i,\text{ref}} }}{N_{\text{atom}} }\right) \right ] ,
&&
\end{flalign}

\noindent where $p^{\left(\alpha\right)}_{i,\text{pred}}$ and $p^{\left(\alpha\right)}_{i,\text{ref}}$ represent the predicted and reference values of the $\alpha$-th Cartesian component of $\mathbf{p}$ for the $i$-th molecule. To evaluate the performance of the model, we predicted the $\mathbf{p}_{\text{ref}}$ of 3\,580 PAH conformations excluded from the training set. As shown in Figure \ref{F2}c, there is strong agreement between the predicted and reference values, with a ${R^2}$ reaching 0.97 and a mean RMSE of 0.0545 debye, as shown in Table \ref{T1}.

\subsection{Anharmonic IR spectrum} 
\label{AIR}
To compute the anharmonic IR absorption spectrum of a PAH, MD simulations were performed to simulate its finite-temperature dynamics based on atomic forces predicted by the trained NNFF. We used the MD implementation in the Atomic Simulation Environment (ASE) \citep{larsen2017}. Initially, the molecule reaches thermal equilibrium in 200 ps in a canonical ensemble (NVT) at a target temperature controlled by the Nosé-Hoover thermostat, with a timestep of 0.5 fs. The system was then further simulated in a microcanonical ensemble (NVE) for 200 ps, recording atomic configurations at every 1.0 fs during vibration.

These configurations serve as input to the EPNN, which is used to compute $\mathbf{p}(t)$ at each time step of the vibration. The dipole time autocorrelation function $\langle \mathbf{p}(0) \cdot \mathbf{p}(t) \rangle$, characterizing the time evolution of $\mathbf{p}$, is then subjected to a Fourier transform to obtain the absorption cross-section $\sigma$ (in electrostatic unit) \citep{berens1981}:
\begin{flalign}
\label{eq4}
&&
\sigma(\omega)  = \frac{2\pi \omega (1 - \text{e}^{-\hbar\omega/k_{\text{B}}T})D(\omega)}{3\hbar c} \int_{0}^{\infty} \text{e}^{-i \omega t} \langle \mathbf{p}(0)\cdot\mathbf{p}(t) \rangle dt,
&&
\end{flalign}

\noindent where $\omega$ denotes the angular frequency, $k_{\text{B}}$ represents the Boltzmann constant, $\hbar$ is the reduced Planck constant, $c$ is the speed of light, $T$ indicates the temperature, $t$ is time, and $D(\omega)$ is the quantum correction factor, which is set to 1.0 in this work. Further explanation of this equation is provided in the Supplementary Information.

$\sigma$ is connected to the laboratory-measured IR intensity $I$ as:
\begin{flalign}
\label{eq4}
&&
\int_\omega \sigma(\omega) d\omega = 2\pi \cdot 10^5 \frac{c \cdot I(\omega)}{N_{A}},
&&
\end{flalign}

\noindent where $N_{A}$ is Avogadro’s number. To compare $I$ with the MLMD computed $\sigma$, we applied Lorentzian broadening to the experimental spectrum with a full width at half maximum (FWHM) of $2 \text{cm}^{-1}$. While the error values vary with the FWHM choice (see Supplementary Information), our key conclusion remains unaffected by this selection.

\subsection{Reference methods for comparison} 
\label{MtC}

To evaluate the accuracy of our model, we compared it with two established QCC methods to calculate the IR spectra of PAHs. The first method, designated as \textit{DFT harmonic}, is a widely adopted hybrid DFT approach for harmonic spectrum calculations. In this method, the vibrational frequencies (normal modes) are calculated from the second derivatives of the potential energy with respect to the positions of the nuclei at the stationary point of the system. IR intensities are determined using the double harmonic approximation, which involves computing the derivatives of $\mathbf{p}$ with respect to the normal modes \citep{langhoff1996}. The calculations were carried out at the B3LYP/4-31G level of theory using Gaussian 16, ensuring consistency with the QCCs employed in our training dataset. It is important to note that the assumption of a harmonic potential frequently underestimates fundamental frequencies, often necessitating empirical scaling factors to align the computational results with the experimental data \citep{bauschlicher1997}. Despite the transferability issues introduced by these empirical scaling factors, this method remains the most commonly utilized method for molecular IR spectrum calculations due to its relatively low computational cost. For the sake of a fair comparison, we do not apply any scaling factors to the DFT harmonic results, as these factors should ideally be derived from fitting to experimental spectra. The DFT harmonic method was used to compute the harmonic spectra for 49 PAH molecules, for which the experimentally measured spectra are available in the PAHdb.

The second method used for comparison is the VPT2 anharmonic spectrum calculations combined with B3LYP/6-311+g* DFT, referred to here as \textit{DFT anharmonic}. This approach represents the state-of-the-art for computing the molecular anharmonic IR spectra, and has been implemented in Gaussian 16. It requires the computation of the second, third, and fourth derivatives of the potential, which results in significant computational demands \citep{nielsen1951,mackie2015}. Due to this cost, we applied VPT2 to calculate the anharmonic spectra of 26 PAHs, with the maximum $N_C$ being 32, from the 49 PAHs used for the DFT harmonic method.

The 49 PAHs for which the experimental spectra are available in PAHdb (v3.2), ranging from 10 to 50 carbon atoms, exhibit a variety of chemical structures, as detailed in the Supplementary Information. The experimental IR spectra were obtained using matrix isolation techniques at low temperatures \citep{hudgins1995, hudgins1998a, hudgins1998b, hudgins1998c, hudgins1999, mattioda2003, mattioda2005, mattioda2014, mattioda2017}. To facilitate comparison with experimental data, we employed theoretical spectra computed using MLMD at an effective temperature of 50 K, which accounts for molecular vibrations under experimental conditions \citep{esposito2024b}. Stochastic uncertainty analysis confirms size-independent prediction stability across PAHs, and the model inherently supports non-planar geometries (see, Supplementary Information).

\section{Results and discussion}

\begin{figure}
\centerline{\includegraphics[width=0.5\textwidth]{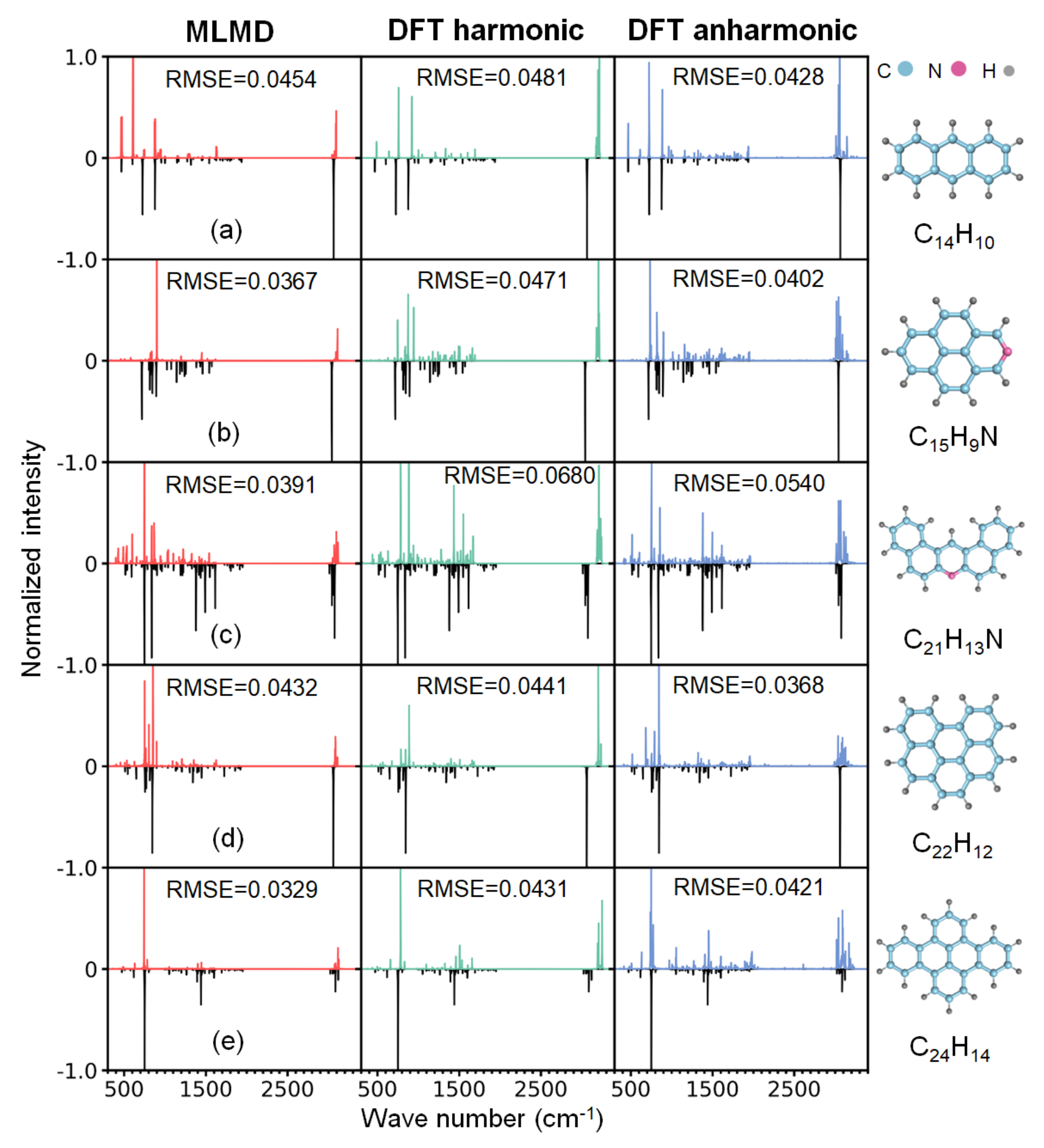}}
\caption{Comparison of IR spectra for five PAHs computed using the MLMD at 50~K(left, red), DFT harmonic (middle, green), and DFT anharmonic (right, blue) methods, alongside the experimentally measured spectra (black). The intensities are normalized to the maximum value in each spectrum.}
\label{F3} 
\end{figure}

To assess the precision of our predictions relative to the experimental data, Figure \ref{F3} presents the computed spectra for five PAHs of increasing sizes, ordered from top to bottom. The black curves represent the experimental spectra, while the red, green, and blue curves show the spectra computed using MLMD, DFT harmonic, and DFT anharmonic methods, respectively. As seen in the figure, the anharmonic spectra (represented by the red and blue curves) exhibit better agreement with the experimental data than the harmonic spectra (green), both in terms of frequency and intensity.

Among the anharmonic spectra, we find that the traditional DFT anharmonic method performs better than the MLMD method for the smallest PAH. However, for larger PAHs, the MLMD method provides more accurate predictions. Specifically, for the molecule \ce{C14H10}, the spectra predicted by DFT anharmonic achieved a RMSE value of 0.0428. In contrast, the RMSE values for MLMD were higher, at 0.0454, while the DFT harmonic method produced even larger errors of 0.0481, owing to a significant mismatch in band positions, as shown in the middle panels of Figure \ref{F3}a. For larger PAHs, such as \ce{C21H13N} and \ce{C24H14}, the MLMD method performs comparably to DFT anharmonic in predicting IR intensities.

\begin{figure}
\centerline{\includegraphics[width=0.48\textwidth]{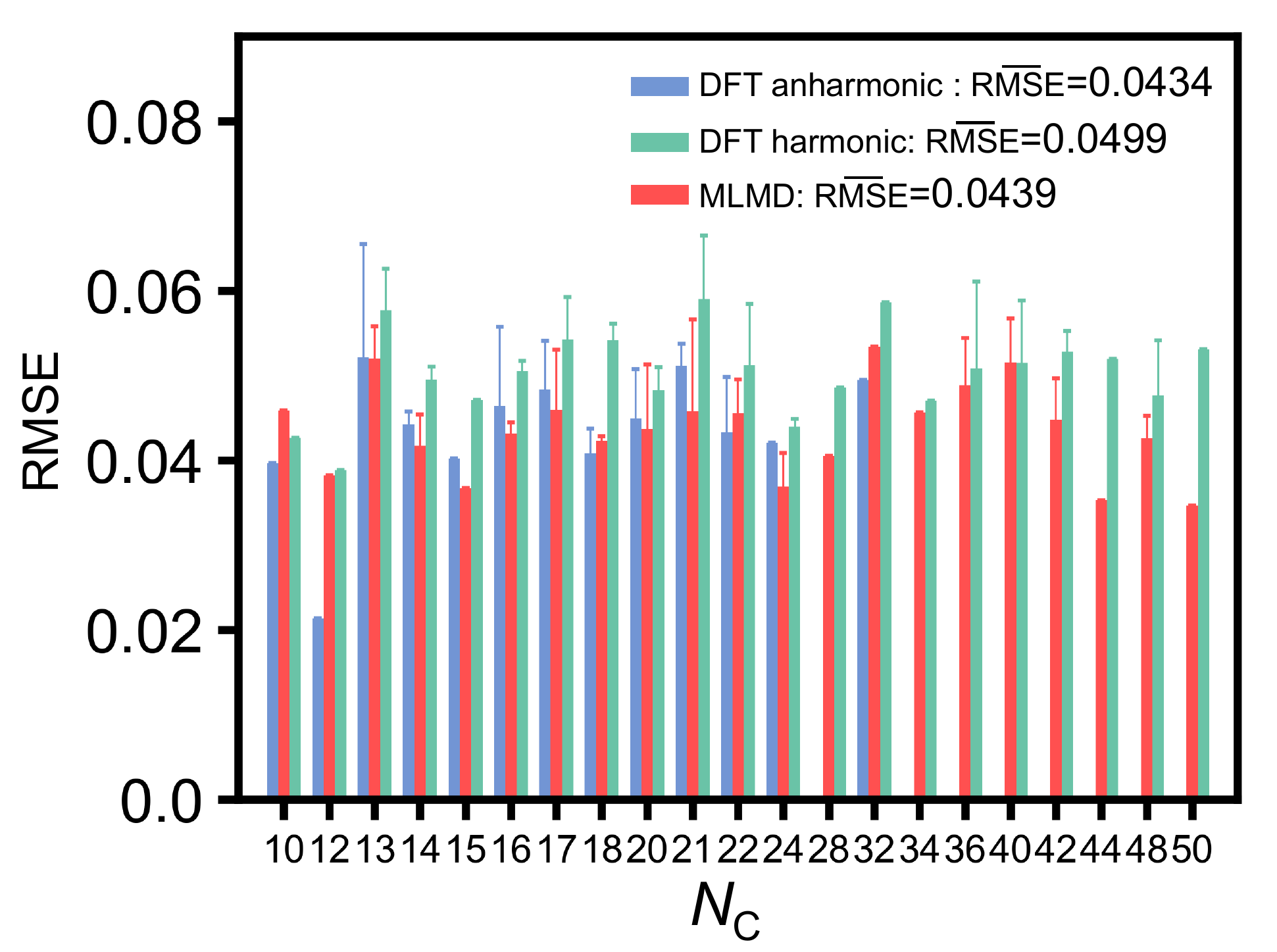}}
\caption{Distribution of RMSE values for spectra predicted by the three computational methods, compared to the experimental spectra of the 49 PAHs. $N_\text{C}$ represents the number of carbon atoms in each molecule. Note that the anharmonic DFT calculation is not performed for large PAHs (Section \ref{MtC}).}
\label{F4}
\end{figure}

Figure \ref{F4} presents an overview of the error distribution in predicting the IR spectra of the 49 PAHs relative to the experimental data, for each of the three computational methods. The results clearly demonstrate that the harmonic spectra exhibit significantly poorer agreement with the experimental data compared to the anharmonic spectra, consistent with the findings of previous studies \citep{mackie2015,maltseva2016,lemmens2019}. The precision of the MLMD and DFT anharmonic methods is comparable, with mean RMSE values of 0.0439 and 0.0434, respectively. For PAHs with fewer than 12 carbon atoms, the DFT anharmonic method generally outperforms MLMD. However, as the size of the PAHs increases, MLMD tends to provide comparable predictions than DFT anharmonic in terms of normalized spectra.

To further clarify the differences between MLMD and DFT in specific vibrational regions, a mode-by-mode error analysis is presented in Table \ref{T2}. The results show that MLMD achieves higher accuracy in the 2.6–3.5, 5.2–5.8, and 6.0–10.0 {\textmu}m regions, corresponding to C–H stretching, combination bands, and C–H in-plane (IP) bending modes, respectively. In contrast, for out-of-plane (OOP) bending modes in the 10.0–15.0 {\textmu}m region, the DFT-based anharmonic method demonstrates better agreement with experimental data.

\begin{figure}
\centerline{\includegraphics[width=0.45\textwidth]{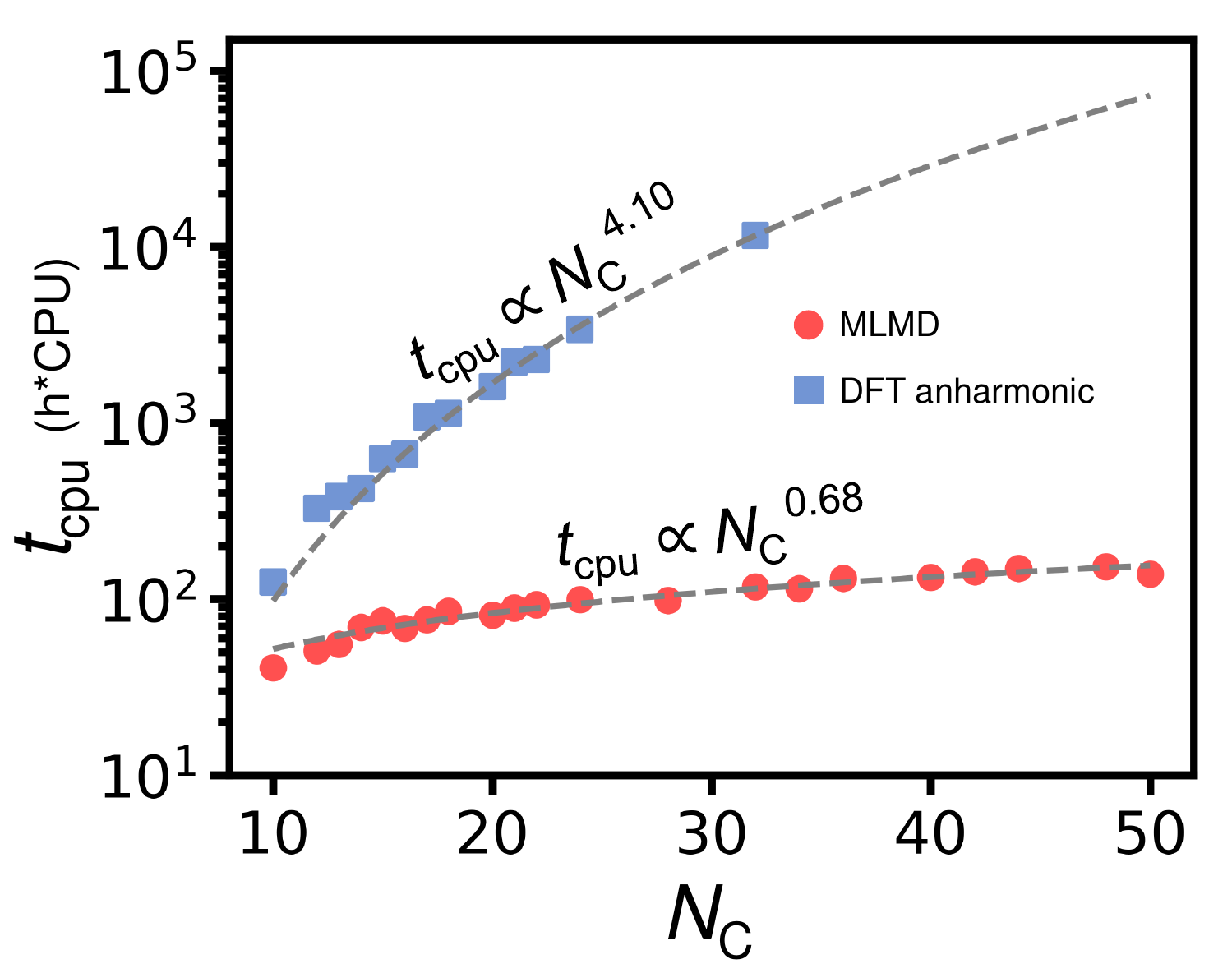}}
\caption{Average CPU time (run time $\times$ number of cores, on a logarithmic scale) required to compute the anharmonic IR spectrum of a PAH as a function of the number of carbon atoms, using the MLMD (red dots) and DFT anharmonic (blue dots) methods. The DFT anharmonic computations were parallelized on a server with 40 cores (Intel Xeon E5-2680 CPU at 2.40 GHz), while MLMD jobs were run sequentially on a CPU of the same model.}
\label{F6}
\end{figure}

Despite having comparable accuracy, the computational efficiency of MLMD is significantly superior to that of the traditional DFT anharmonic method. Figure \ref{F6} shows the average CPU time, $t_\text{cpu}$, required to compute an anharmonic IR spectrum as a function of the molecular size. It is evident that the $t_\text{cpu}$ for MLMD is initially approximately five times shorter than that for the DFT anharmonic for the smallest PAHs. As the molecular size increases, the difference in $t_\text{cpu}$ grows, eventually reaching nearly two orders of magnitude for the largest PAH in the set of 49 molecules.

\begin{table}
\centering
{\caption{\label{T2} Mean RMSE values of IR spectra predicted by MLMD and DFT anharmonic methods, with respect to experimental spectra of the 26 PAHs across spectral regions of key vibrational modes.}}
\begin{tabular}{>{\centering\arraybackslash} lccr}
\hline
Wavelength ({\textmu}m) & Vibration mode & DFT anharm. & MLMD\\
\hline
2.6-3.5 & CH stretching & 0.0534 & 0.0403\\
5.2-5.8 & Combination   & 0.0103 & 0.0044\\
6-10  & CH IP bending  & 0.0301 & 0.0273 \\
10-15 & CH OOP bending  & 0.0892 & 0.1067\\
\hline
\end{tabular}
\end{table}

The trend of increasing CPU time with molecular size for MLMD is remarkably different from that of the DFT anharmonic method. To provide a quantitative illustration, a simple power-law fit of type $y=ax^{b}$ was applied to the data points in Figure \ref{F6}, as indicated by the dashed curves. These best-fitting curves show that the DFT computation of anharmonic spectra for large PAHs becomes extremely time-consuming, with a computational scaling of $\sim N_\text{C}^{4}$ for the DFT anharmonic method. In contrast, MLMD demonstrates a significantly more efficient scaling of approximately $N_\text{C}^{0.7}$. This results in a computational time difference of approximately three orders of magnitude for a PAH with $N_\text{C}=50$. This enhanced efficiency makes MLMD well suited for high-throughput computation of molecular anharmonic spectra, particularly for large PAHs, which are considered key sources of AIBs, and for comprehensively studying the spectroscopic properties of PAHs with extensive structural diversity.

The AIB features are believed to originate from the IR emission of PAHs excited by UV photons. Upon absorbing such a photon, a PAH becomes highly energized, with its effective temperature potentially rising to several hundred or even over a thousand kelvin. Consequently, the low-temperature spectra of PAHs mentioned earlier have limited astronomical relevance in this context. Traditionally, the Wang and Landau method has been used to incorporate temperature effects into a DFT-computed IR spectrum \citep{Wang2001, basire2009}. However, as discussed above, such DFT calculations are computationally expensive. Moreover, the Wang-Landau algorithm often requires extensive sampling to ensure convergence, making it suitable mainly for studying individual, very small PAHs \citep{chen2018a, chen2018b, mackie2018b, mackie2021, chakraborty2021}. In contrast, the MLMD approach employed in this work explicitly incorporates temperature effects, providing a more efficient solution due to its high computational efficiency and accuracy.

\begin{figure}
\centerline{\includegraphics[width=9cm]{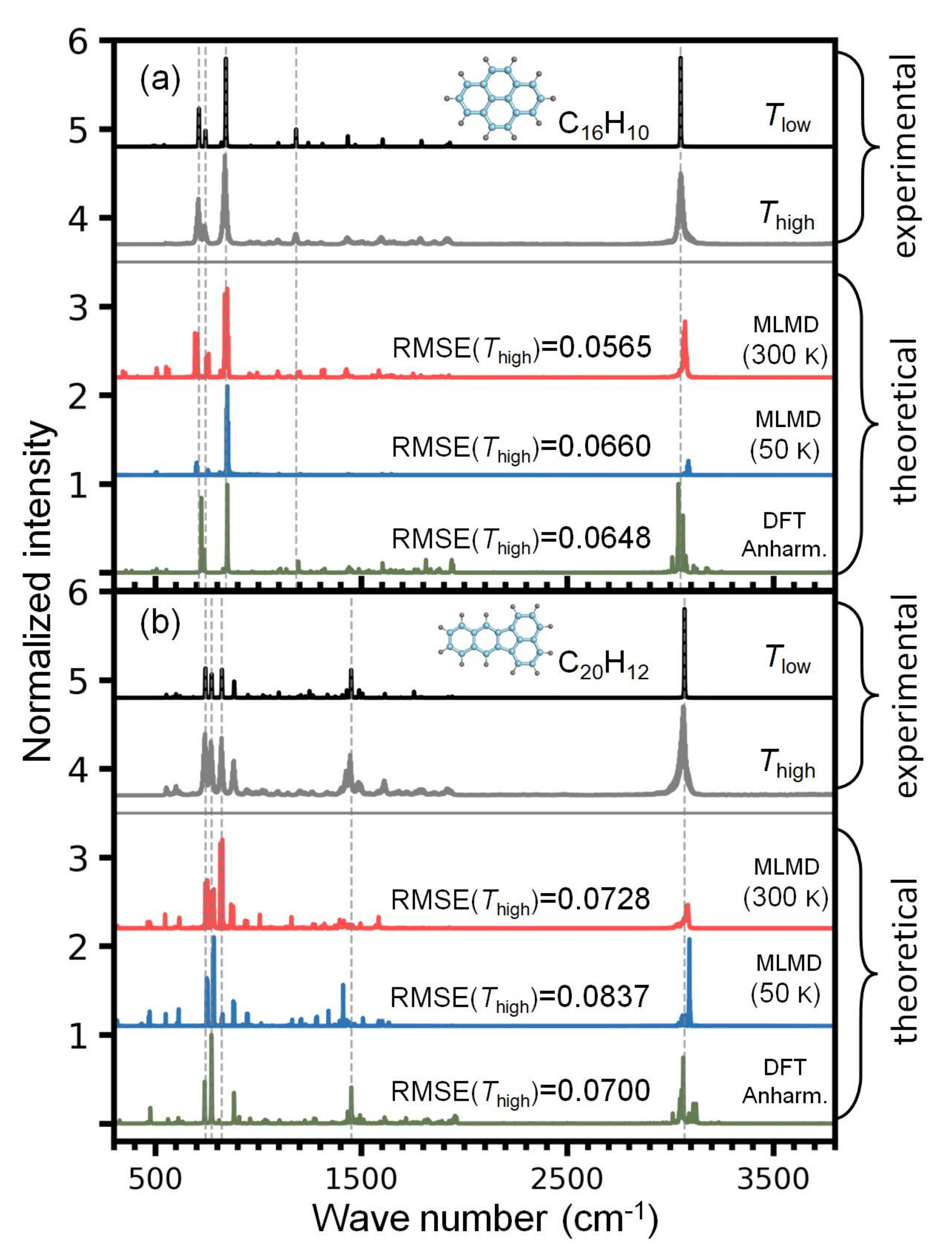}}
\caption{Comparison of IR spectra for pyrene \ce{C16H10} (a) and benzo[$k$]fluoranthene \ce{C20H12} (b) between experimental data (top panels) and theoretical predictions by MLMD at two different temperatures and by the DFT anharmonic method (bottom panels).}
\label{F7}
\end{figure}

To demonstrate the precision, we compared theoretical anharmonic spectra with experimental IR spectra of two PAHs, pyrene and benzo[$k$]fluoranthene, measured at different temperatures. The higher temperature gas-phase experimental spectra have been taken from the NIST Chemistry WebBook database \citep{Linstrom2001}, while the lower temperature spectra come from matrix isolation experiments of \citet{hudgins1998a,hudgins1999}. We note that the gas-phase IR spectra do not include explicitly stated temperature values, but prior studies (e.g. \citet{mackie2015,mackie2016}) assumed an approximate measurement temperature of about 300 K. Figure \ref{F7} shows the NIST spectra (gray curves), referred to as $T_\text{high}$, and the low-temperature spectra (black curves), referred to as $T_\text{low}$. Comparison of the two sets of experimental spectra measured at different temperatures reveals that thermal effects lead to notable changes in spectral characteristics, such as band broadening and peak shifts (indicated by dashed lines), which aligns with previous observations of \citet{chen2018a}.

The two pairs of experimental spectra shown in the top panels of Figure \ref{F7} a and b are compared with three theoretical spectra in the bottom panels, calculated using MLMD at 300 (red) and 50 K (blue), as well as by the anharmonic DFT method (green). Compared with the 50 K MLMD results, the temperature-induced band broadening and peak shift behaviors are better captured by the MLMD results at 300 K, with relatively low RMSE values of 0.0565 and 0.0728 for the two molecules. In contrast, the agreement with the gas-phase experimental spectra is poorer for the 50 K results.

Despite the growing need to incorporate anharmonicity and temperature effects into astronomical spectroscopic analyses, driven by high-resolution infrared observations in the JWST era \citep{peeters2021}, only a limited number of theoretical anharmonic spectra for PAHs, likely fewer than 100, are available in the literature. Typically, these studies focus on just 2-5 PAH species, most of which contain fewer than 20 carbon atoms \citep{esposito2024c, esposito2024d, mackie2022, mackie2021}. This narrow scope makes it difficult to draw general conclusions, especially given the vast diversity of PAH chemical structures. Such limitations hinder the accurate interpretation of AIBs and are mainly due to the high computational cost of traditional QCCs. To address this gap, we have computed the anharmonic spectra at 50, 300, and 600 K for 1,704 molecules from version 3.20 of PAHdb, which includes neutral PAHs as large as 216 carbon atoms. This dataset is open to the public as described in the Section of Data Availability. 

\begin{figure}
\centerline{\includegraphics[width=0.48\textwidth]{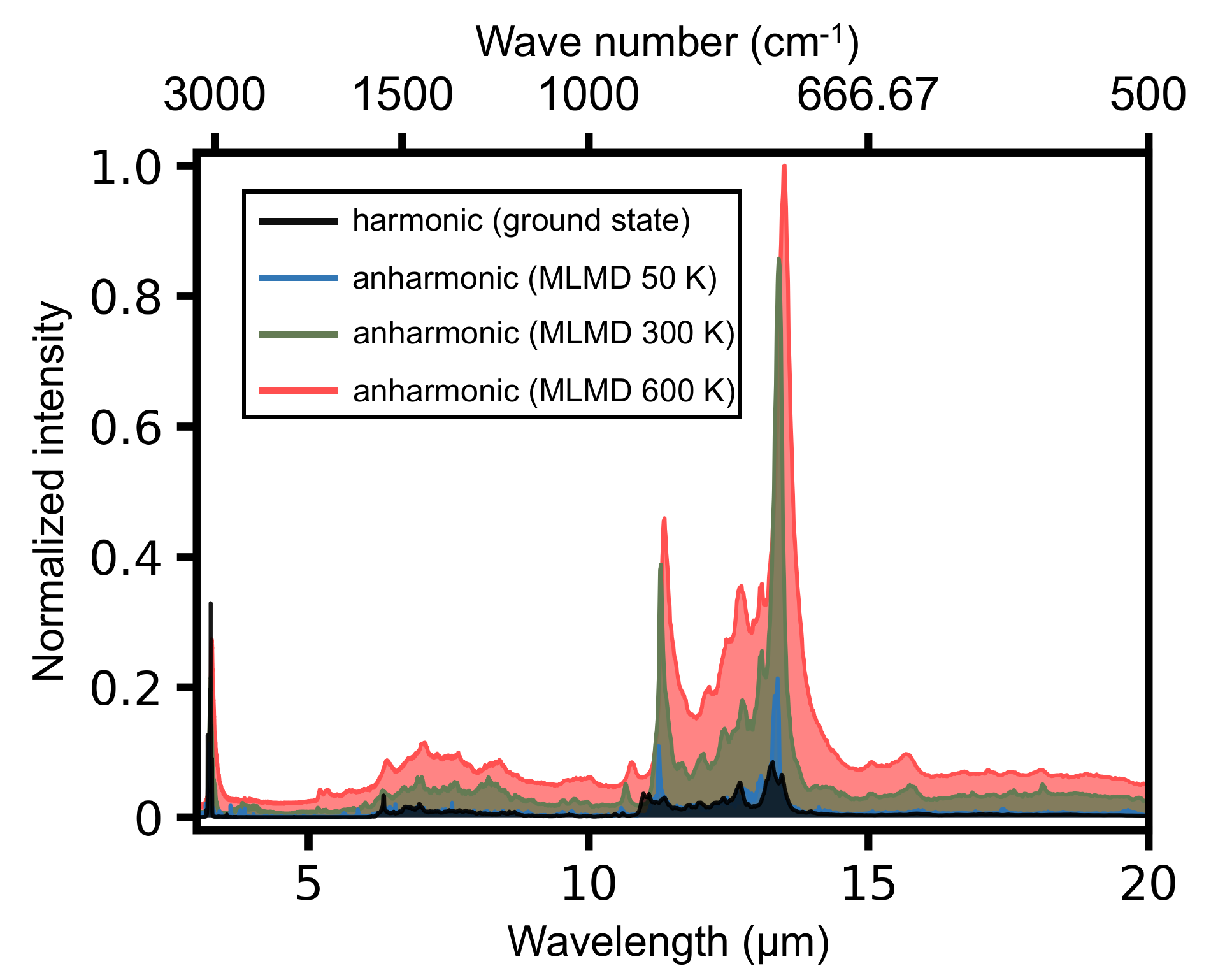}}
\caption{Absorption spectra of a mixture of 1,704 neutral PAHs computed using MLMD at 50 K (blue), 300 K (green), and 600 K (red). The three MLMD spectra were first normalized with respect to the total intensity of the 50 K spectrum, while the scaled harmonic spectrum from PAHdb (black) was normalized to its own total intensity.}
\label{F8}
\end{figure}

This dataset is expected to be a valuable resource for the spectral decomposition of AIBs and for data-driven studies investigating the precise structure-spectral relationships of PAHs. In addition, it will aid in the exploration of temperature effects. For instance, Figure \ref{F8} displays the anharmonic IR spectra of a simple mixture of the 1\,704 neutral PAHs at 50 K (blue), 300 K (green), and 600 K (red). These spectra represent the sum of individual PAH spectra and are compared to the ground-state harmonic spectra from PAHdb (black, with scaling factors). Notably, the IR signals are significantly enhanced at high temperatures, exhibiting stronger peaks and an elevated plateau. A redshift in peak positions is also observed at high temperatures as a general trend. Furthermore, some emission bands that are weak at lower temperatures become significantly stronger at 300 and 600 K. For example, a distinct band emerges in the 3.5–6 $\upmu$m region, absent in most harmonic spectra but observed in AIBs \citep{jourdain1986,sloan1997}. This feature may originate from the redshift of the C–H stretching band in superhydrogenated PAHs due to anharmonic effects \citep{mackie2018a,Yang2020,Sandford2013}, with additional redshift and intensity enhancement likely driven by temperature effects.

\section{Conclusions} 
We have demonstrated that, when applied to the prediction of the IR spectra of PAHs, MLMD provides predictive accuracy comparable to that of traditional quantum chemical methods but at a fraction of the computational cost. With a scaling law of approximately $N_\text{C}^{0.7}$, MLMD is particularly suitable for large PAHs, which are key contributors to AIBs. In contrast, traditional quantum chemical anharmonic methods scale much more steeply, approximately $N_\text{C}^{4.1}$. Furthermore, we have shown that MLMD is an ideal method for studying temperature effects in molecular IR spectra, which might be critical for understanding astronomical observations. By calculating the anharmonic spectra of 1,704 neutral PAHs from PAHdb at 50, 300, and 600 K, with molecular sizes containing up to 216 carbon atoms, we highlight that the efficiency of MLMD enables the creation of extensive molecular anharmonic spectral datasets. These datasets could be of instrumental importance for the advancement of AI-assisted astronomical analyses of observational IR spectra in the future.

Despite the impressive efficiency of MLMD, several limitations persist. First, the current model was trained exclusively on neutral PAHs with natural chemical elements and does not account for charged molecules or isotopologues, both of which may play a significant role in astronomical contexts. This limitation stems from the scarcity of data for the IR spectra, as well as challenges in incorporating charge state and isotope information into molecular descriptors. Second, because of the data-driven nature of the approach, the model may exhibit increased uncertainty when predicting spectra for molecules that deviate substantially from those in the training dataset. While MLMD captures a broad spectrum of anharmonic effects through classical sampling of the potential energy surface, it does not account for quantum-specific phenomena such as Fermi resonance, which arise from interactions between quantized vibrational states and require a fully quantum mechanical treatment. Third, the training data of the current model were mainly sourced from molecular conformations at 300 K, which are in principle not sufficient for reliably capturing IR spectral features at very high temperatures (e.g., >600 K). We are actively working to address these challenges, with the goal of further enhancing MLMD's capabilities for IR spectrum computation.

\section*{Supplementary Materials}
The source code and the MLMD model to calculate the anharmonic spectra of PAHs using MLMD are freely available at \href{https://doi.org/10.5281/zenodo.14998197}{DOI: 10.5281/zenodo.14998197}. The model is ready for use without the need for further ML training. It will be continuously updated to improve predictive performance. This repository also includes the spectral data for the 1\,704 theoretically-calculated and 49 experimentally-tested PAHs. The Supplementary Information is also included in this repository.


\bsp
\label{lastpage}
\end{document}